\documentclass[runningheads]{llncs}

\usepackage{amsmath}
\usepackage{subcaption}
\usepackage{booktabs}
\usepackage{graphicx}
\usepackage[misc]{ifsym}
\usepackage[colorlinks,allcolors=blue]{hyperref} 

\usepackage{pdfpages}
\usepackage{tikz}
\usepackage{pgfplotstable}
\usepackage{pgfplots}
\usepackage{pdfpages}

\begin{document}
\title{SPAN: Subgraph Prediction Attention Network for Dynamic Graphs}

\author{
    Yuan Li \and
    Chuanchang Chen \and
    Yubo Tao\textsuperscript{(\Letter)} \and 
    Hai Lin\textsuperscript{(\Letter)}
}

\authorrunning{Y. Li et al.}

\institute{
    State Key Lab of CAD\&CG, ZheJiang University, HangZhou, China \\
    \email{\{yuanli, chenchuanchang\}@zju.edu.cn} \\
    \email{\{taoyubo, linhai\}@cad.zju.edu.cn} 
}
\maketitle              
\begin{abstract}
    This paper proposes a novel model for predicting subgraphs in dynamic graphs, an extension of traditional link prediction. This proposed end-to-end model learns a mapping from the subgraph structures in the current snapshot to the subgraph structures in the next snapshot directly, i.e., edge existence among multiple nodes in the subgraph.  A new mechanism named cross-attention with a twin-tower module is designed to integrate node attribute information and topology information collaboratively for learning subgraph evolution. We compare our model with several state-of-the-art methods for subgraph prediction and subgraph pattern prediction in multiple real-world homogeneous and heterogeneous dynamic graphs, respectively. Experimental results demonstrate that our model outperforms other models in these two tasks, with a gain increase from $5.02\%$ to $10.88\%$.
    
\end{abstract}

\section{Introduction}
 \label{sec:introduction}
 An essential part of network analysis is network evolution analysis~\cite{network_evolving_begin,Evolving_network,evolving_topology}, especially subgraph evolution analysis, such as the purchase intention of a group of users in user-product networks. However, previous subgraph research studies~\cite{subgraphnn11,subgraphnn12,subgraphnn13} rarely focus on the subgraph prediction problem in subgraph evolution analysis: predicting future connectivity within a subgraph in dynamic graphs.

 An intuitive approach for subgraph prediction is a two-stage scheme, including a node embedding method and traditional link prediction. Firstly, a node embedding method~\cite{HGT,egcn} generates low-dimensional vector representations of nodes. Then, the edge existence of  $\frac{k(k-1)}{2}$ edges in a $k$-node subgraph is independently predictable through traditional link prediction.

 However, there are two limitations to this approach. First, traditional link prediction typically requires users to specify or learn a global threshold from data to determine the existence of edges rather than adaptively adjusting the threshold for different local subgraphs. Second, higher-order structures~\cite{higher_order_link_prediction,linkdependency}, such as network function blocks~\cite{network_functional_block}, are ignored. Due to higher-order structures, edges in networks are not independent, i.e., an edge's establishment or disappearance may depend on both the similarity between two nodes and their adjacent edges~\cite{linkdependency}. Subgraph pattern neural networks (SPNN)~\cite{SPNN} uses a joint prediction mechanism to solve the edge dependency limitation. However, SPNN focuses on the subgraph pattern prediction problem, which requires both subgraph patterns predefined by humans and subgraphs with a fixed size. Therefore, we design a new method for the subgraph prediction to solve both limitations.

 Besides the global threshold and edge dependency limitations, previous research studies on subgraphs predefine~\cite{SPNN} or ignore~\cite{subgraphnn11,subgraphnn12,subgraphnn13} the collaborative relationship between node attribute information and topology information, e.g.,  extracting important information from node attributes and topologies separately, and concatenating their representations at last~\cite{subgraphnn11}. However, important topology information may also be based on node attribute information, and critical node attribute information may also be related to topology information, which means we cannot deal with them separately. For instance, many topology structures in subgraphs are more important than other topology structures for subgraph evolution due to their specific node attributes, such as meta-path~\cite{hemoimportant} and local structures with high-connection nodes~\cite{degree}. Similarly, nodes with higher topology centrality in subgraphs are more critical than other nodes in subgraph evolution~\cite{center}. Although previous research studies can predefine the collaborative relationship artificially, the real collaborative relationship changes with graphs and only a part of this relationship (e.g., only the relationship between 3-nodes) is covered by the predefined relationship. Therefore, we propose a new method that integrates node attribute information and topology information to extract essential data features for subgraph evolution without human participation.

 This paper proposes a novel end-to-end \textbf{S}ubgraph \textbf{P}rediction \textbf{A}ttention \textbf{N}etwork (SPAN)  model to learn a mapping from subgraph structures in the current snapshot to subgraph structures in the next snapshot. For the global threshold limitation, we introduce an end-to-end learning mechanism to avoid the global threshold. We also use a joint prediction mechanism~\cite{SPNN,higher_order_link_prediction} to solve the edge dependency limitation. Furthermore, we develop a twin-tower module with the cross-attention mechanism to extract important data features for subgraph evolution by considering the collaborative relationship between node attribute information and topology information. Our main contributions are summarized as follows:

 \begin{itemize}
  \item We propose a new model, named SPAN, for subgraph prediction and a variant named SPAN-H for subgraph pattern prediction. To the best of our knowledge, SPAN is the first end-to-end model designed to predict the evolution of arbitrary size subgraphs.
  \item We propose a new mechanism named cross-attention with a twin-tower module for solving the collaborative limitation.
  \item Experimental results demonstrate that our method is more effective and scalable than state-of-the-art subgraph prediction and subgraph pattern prediction methods. 
 \end{itemize}

 \section{Related Work}
 \label{sec:rel}

 Previous methods based on node embeddings and traditional link prediction can be divided into static graph embedding methods and dynamic graph embedding methods. Static graph embedding methods~\cite{GCN,graphsage,GAT,HAN,HGT} are not designed for dynamic graphs, predicting the future connectivity of subgraphs solely based on the current snapshot. In contrast, dynamic graph embedding methods~\cite{CTDNE,TNE,triad} predict subgraphs' future connectivity based on current and previous snapshots. Recently, with the massive success of the attention mechanism in temporal information extraction, some attention-based dynamic graph embedding methods have been proposed~\cite{dysat,egcn}.

 Dynamic subgraph prediction methods focus on the dynamic evolution of subgraphs, which involves higher-order structures with multiple nodes. Previous dynamic subgraph prediction methods are limited, such as edge dependency limitation and global threshold limitation. Higher-order link prediction~\cite{higher_order_link_prediction} and SPNN~\cite{SPNN} overcome the edge dependency limitation by jointly predicting the connection between multiple nodes. Nevertheless, these methods still have some restrictions, such as human-predefined fixed-sized subgraphs and subgraph evolution patterns. Compared with previous methods, our method overcomes these two limitations and removes these restrictions using a more powerful model and new mechanisms. We also identify a new collaborative limitation that the collaborative relationship between node attribute information and topology information on subgraph evolution, which has been ignored or predefined by humans in previous subgraph research. We address this new limitation using a new mechanism named cross-attention.

 \section{Proposed Method}
 
 Our method is composed of a Bayesian subgraph sampling algorithm and SPAN. Bayesian subgraph sampling is responsible for generating subgraphs, and SPAN model is used for the subgraph prediction of dynamic graphs.
 
 \subsection{Bayesian Subgraph Sampling}
 \label{sec:BSSM}

 Given a dynamic graph  $\Gamma$ with T snapshots $\{G_1,...,G_T\}$,  $G_t=(V_t,E_t)$ is a continuous-time graph (i.e., one snapshot),  $V_t$ is a node set with node attributes and $E_t$ is an edge set with weights. A subgraph $S_t = (V_t^s, E_t^s)$ is a subset of the snapshot $G_t$, such that $V_t^s \subseteq V_t$ and $E_t^s = \{(u,v) | u \in V_t^s, v \in V_t^s$, and $(u, v) \in E_t \}$. Bayesian subgraph sampling aims to sample a series of subgraph evolution pairs $(S_t,S_{t+1}),...$ from $\Gamma$.

First, we randomly choose a snapshot $G_t=(V_t,E_t)$.
Second, we sample a node $v$ from $G_t$ randomly, initialize the subgraph $V_t^s = \{v\}$ and determine the number of nodes $n$ sampled from a uniform distribution between 3 and $k$ (the maximum size of subgraphs). 
Third, we randomly select a node $v_i$ from $V_t^s$ with the probability $1-\alpha$ for connected subgraphs or randomly select a node $v_j$ in $V_t/V_t^s$ with the probability $\alpha$ ($\alpha=0.01$) for disconnected subgraphs. For $v_i$, we randomly select an adjacent node $v_j$ with the transfer probability $p(v_j|v_i)$ to expand the subgraph.  Motivated by the Bayesian network theory, the transfer probability $p(v_j|v_i)$ can be computed as
\begin{figure*}
  \centering
  \includegraphics [width=\textwidth]{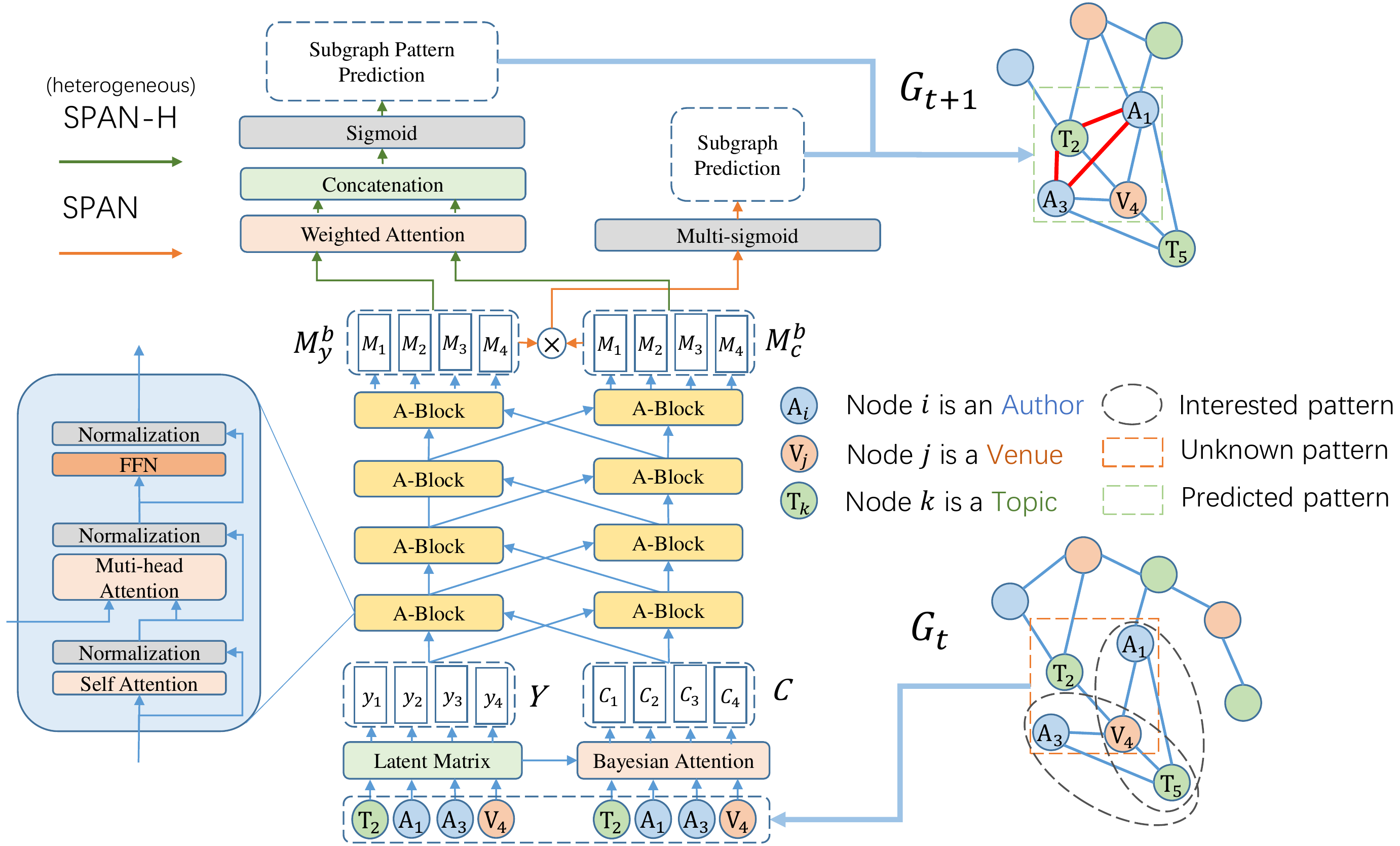}
  \caption{The architecture of SPAN/SPAH-H with one subgraph as an example.}
  \label{fig:model}
  \end{figure*}
\begin{equation}
  p(v_j|v_i)=\frac{w_{j,i}}{\sum_{(v_k,v_i)\in E_t}w_{k,i}},
\end{equation}
where $w_{j,i}$ is the weight of an edge $(v_j,v_i)$ for weighted graphs, and $w_{j,i} = 1$ for unweighted graphs.
For $v_j$, we add this node to the subgraph as a random jump operation. 
Repeat this node expansion process until the number of nodes in $V_t^s$ reaches $n$ to generate a subgraph. Finally, according to the node set $V_t^s$, we sample $S_t$ from the current snapshot $G_t$. 
$S_{t+1}$ can be constructed via $V_{t+1}^s = V_t^s$ and $E_{t+1}^s = \{(u, v) | u \in V_{t+1}^s,  v \in V_{t+1}^s$,  and $(u, v) \in E_{t+1}\}$ to compose an evolution pair of subgraphs $(S_t,S_{t+1})$. 

Repeating Bayesian subgraph sampling, we obtain sufficient connected/disconnected subgraph evolution pairs. 
In addition, the Bayesian network theory minimizes the probability that nodes with many links to the subgraph (i.e., nodes having significant influence over the subgraph) are not included in the subgraph.
 \subsection{Subgraph Prediction}
 \label{sec:EM}
 Given a $k$-node subgraph $S_t = (V_t^s, E_t^s)$ in the snapshot $G_t$ ($t \in [1, T-1]$), the goal of subgraph prediction is to predict the structure of the subgraph $S_{t+1} = (V_{t+1}^s, E_{t+1}^s)$ in $G_{t+1}$, where $V_t^s = V_{t+1}^s$. Therefore, we propose  an end-to-end subgraph pattern/structure prediction model named SPAN. 

 As shown in Figure~\ref{fig:model}, the architecture of SPAN has two inputs $Y\in R^{k\times D}$ and $C\in R^{k\times D}$, where $D$ is the dimension of the embedding space. For each pair of subgraphs $(S_t, S_{t+1})$,
 we consider the structure of $S_t$ and $S_{t+1}$ as the source and target language, respectively. Thus, we flatten the subgraph as a node sequence using the sampling order of nodes during Bayesian subgraph sampling. Each node can be assigned a dense embedding vector to encode its information. All embedding vectors of the entire graph are stored in a latent matrix. The first input $Y$ of SPAN is the nodes' self-information in the subgraph by extracting the corresponding embedding vectors $y$s from the latent matrix and constructing a set $Y=(y_1,...,y_k)$. In detail, the latent matrix represents the graph's node attribute information, $Y$ represents the subgraph's node attribute information, and  $y_i$ represents the node embedding vector of the $i$th node in the subgraph, which can be initialized with the degree information of this node in the current snapshot.
 If the number of nodes $n$ in $S_t$ is less than $k$, we can use zero padding for $y_i$, $i \in [n+1,k]$. The second input $C$ of SPAN is the nodes' context information, generated by combining a Bayesian attention layer with the latent matrix to generate inner attention representations $C=(c_1,...,c_k) = PY$, where $P \in R^{k\times k}$ is the Bayesian attention matrix. Intuitively, the second input represents the topological information of the subgraph. 
 We combine all paths (topology information) from node $v_i$ to node $v_j$ in $S_t$ as a Bayesian network $S_{i,j}$; thus, we can compute the joint probability of $S_{i,j}$ as the attention score of node $i$ to node $j$ as follows:
 \begin{equation}
 P_{i,j}=\prod_{v_k\in S_{i,j}}p(v_k|parents(v_k)).
 \label{equ:pij}
 \end{equation}
 Based on two inputs $Y$ and $C$, we design a twin-tower module to synchronously encode the global node attribute information and topology information to generate the intermediate representations $M_y^b$ and $M_c^b \in R^{k\times D}$. Each tower is stacked by $b$ identical attention-based blocks (A-Block). The output of each layer is the input of the next layer. Based on $M_y^b$ and $M_c^b$, we design different prediction modules for subgraph prediction (SPAN) and subgraph pattern prediction (SPAN-H).
 
 \textbf{Attention-based Block}.
 Each attention-based block has three layers: a self-attention layer, a cross-attention layer and a feed-forward layer. A simple self-attention (SA) operation has an input $Y$  representing query, key and value as
 \begin{equation}
 \label{equ:Attention_Y} \ SA(Y) =softmax(\frac{YY^T}{\sqrt{D}})Y.
 \end{equation}
 However, it is difficult for the self-attention mechanism to integrate all dimensional information about queries, keys, and values.
 Therefore, motivated by the transformer structure~\cite{transformer}, we apply the  multi-head attention mechanism to our model for integrating all information.
 Multi-head attention (MA)~\cite{transformer}, composed of $h$ linear projection modules with the same structure, enables the model to focus on different representation subspaces at different positions. Also, because of the collaborative relationship between node attribute information and topology information, they cannot be treated separately.
  Node attribute information can facilitate the extraction of essential topology information,  and topology information can help extract essential node attribute information for subgraph prediction. For example, nodes with high-connection attributes (high-degree nodes) play a more important role than nodes with low-connections, i.e., local topology structures with high-connection attributes in subgraphs have faster evolution speed worth more attention~\cite{degree}. Nodes with similar attributes may have different effects on subgraph evolution due to different subgraph topologies. For instance, nodes with high topology centrality in subgraphs usually have a stronger influence than other nodes on subgraph evolution~\cite{center}. 
  If we extract essential topology information only based on topologies for subgraph evolution, data features of some important topology structures (e.g., meta-path, local structures with high-connection nodes) will be unidentified and lost. The same thing happens with existing methods that extract important node attribute information  for subgraph evolution only based on node attribute information. Similarly, as the complex collaborative relationship between node attribute and topology information varies with graphs and node number of subgraphs, it could not be predefined fully by humans. 
  For example, according to the collaborative relationship, SPNN~\cite{SPNN} defines some important structures artificially and pays more attention to the data features of these structures, but these predefined structures are incomplete, e.g., excluding the local structures with more than four nodes and high-connection nodes. 
  Therefore, we propose a variant of MA named cross-attention to extract important data features based on topology information and node attribute information and include it in A-Block. The formula of the attention block with the cross-attention layer can be described as follows.
  
  First, the formula of MA is
   \begin{equation}
    \label{equ:MultiHead} \ MA(Y)=Concat(head_1,...,head_h)W^o, 
   \end{equation}
   \begin{equation}
    \label{equ:head} \ head_i=Attention(YW_i^Q,YW_i^K,YW_i^V),
   \end{equation}
 where $W^o\in R^{hD\times D}$, $W_i^Q\in R^{D\times D}$, $W_i^K\in R^{D\times D}$ and $W_i^V\in R^{D\times D}$ are the linear projection weight matrices.
To approximate a more complex similarity function, each multi-head attention layer is concatenated with a feed-forward layer, which has the same dimension of input and output with the multi-head attention layer as follows:
  \begin{equation}
  \label{equ:FFN} \ FFN(x) = Max(0, xW_1+b_1)W_2+b_2.
  \end{equation}
Thus, each attention block with the cross-attention layer in the left tower is
 \begin{equation}
 \label{equ:M_e} \ M_y^{i+1} = FFN(MA(M_c^i, M_c^i, SA(M_y^i))),
 \end{equation}
 where $M_c^i \in R^{k\times D}$ is the $i$-th output from the other tower, $M_c^0 = C$ and $M_y^0 = Y$. Similarly, each block in the right tower is 
 \begin{equation}
 \label{equ:M_d} \ M_c^{i+1} = FFN(MA(M_y^i, M_y^i, SA(M_c^i))).
 \end{equation}
 To improve convergence, we add normalization and residual sum after each layer in the attention-based block.
 
 \textbf{SPAN}. We use multi-sigmoid as the final activation function to predict existence probabilities of edges in the subgraph as follows:
 \begin{align}
 \label{equ:M_f_1} \ M_f^{S_t} &= MultiSigmoid(M_y^b(M_c^b)^T).
 \end{align}
 Since this paper focuses on subgraph prediction of undirected graphs, the prediction matrix, i.e., the adjacency matrix, should be symmetric. We enforce this symmetry by the average of the prediction matrix and its transposition as  $M_f = \frac{M_f^{S_t}+{M_f^{S_t}}^T}{2}$.
 Finally, we employ the cross entropy as the loss function as follows:
 \begin{equation}
 \label{equ:loss_sp} \ L =\sum_{S_t\in\hat{S_t}}\sum_{i,j} A_{i,j}^{S_{t+1}}logM^{S_t}_{f_{i,j}}+(1-A^{S_{t+1}}_{{i,j}})log(1-M^{S_t}_{f_{i,j}}),
 \end{equation}
 where $A^{S_{t+1}}\in R^{k\times k}$ is the adjacency matrix of $S_{t+1}$ and $\hat{S_t}$ is the set of sampled subgraphs in the current snapshot.

 \textbf{SPAN-H}. 
 We develop SPAN-H for subgraph pattern prediction in heterogeneous networks. Heterogeneous networks have more abundant node attributes, and subgraph pattern prediction is not the same as subgraph prediction. Therefore, we make two modifications for SPAN. First, we increase the dimensions of the original node embedding vectors to encode abundant node attribute information. Second, subgraph pattern prediction requires embedding vectors for subgraphs. Therefore, we employ a weighted attention layer and a concatenation operation to replace the multi-sigmoid layer in SPAN, as shown in Figure~\ref{fig:model}. The weighted attention layer can be computed by 
 \begin{equation}
 \label{equ:we} \ M_{wy}^{S_t}=\sum_{v\in S_t}\alpha_vM_{y_v}^b,
 \end{equation}
 \begin{equation}
 \label{equ:wd} \ M_{wc}^{S_t}=\sum_{v\in S_t}\beta_vM_{c_v}^b,
 \end{equation}
   where $M_{y_v}^b$ and $M_{c_v}^b$ are the attention information of node $v$ in the output of the last A-Block, as shown in Figure~\ref{fig:model}. 
 The attention score is computed by
 \begin{equation}
 \label{equ:alpha} \ \alpha_v=\frac{e^{\langle M_{y_v}^b,\sum_{u\in S_t}M_{c_u}^b\rangle}}{\sum_{r\in S_t}e^{\langle M_{y_r}^b,\sum_{u\in S_t}M_{c_u}^b\rangle}},
 \end{equation}
 \begin{equation}
 \label{equ:beta} \ \beta_v=\frac{e^{\langle M_{c_v}^b,\sum_{u\in S_t}M_{y_u}^b\rangle}}{\sum_{r\in S_t}e^{\langle M_{c_r}^b,\sum_{u\in S_t}M_{y_u}^b\rangle}}.
 \end{equation}
 Then, the final output is $M_h^{S_t}=Sigmoid([M_{wy}^{S_t}|M_{wc}^{S_t}]W^s)$,
 \begin{equation}
    \label{equ:loss_spp} \ L = \sum_{S_t\in \hat{S_t}}B^{S_{t+1}} logM^{S_t}_h+(1-B^{S_{t+1}})log(1-M^{S_t}_h),
    \end{equation}
    \begin{table}
       \center
         \caption{Statistics of dynamic graphs. $|V|$ = number of nodes; $|E|$ = number of temporal edges; $|T|$ = number of days.}
         \label{tab:datasets}
         \begin{tabular}{rccc}
           \toprule
           Dataset&V&E&T\\ 
           \midrule
           ia-facebook &46,952 &876,993 &1,591\\
           soc-epinions &131,828 &841,372 &944\\
           sx-askubuntu &159,316 &964,437 &2,047\\
           sx-superuser &194,085 &1,443,339 &2,426\\
           wiki-talk &1,140,149 &7,833,140 &2,268\\
         \bottomrule
       \end{tabular}
       \end{table}
 where $W^s\in R^{2D\times 1}$ is the linear projection weight matrix as a classifier, and $M_h\in [0,1]$ is the probability of the predefined subgraph pattern.
 Thus, the loss function becomes
 where $B^{S_{t+1}}\in\{0,1\}$ is a binary value for indicating whether the subgraph pattern exists in the next snapshot. In summary, we introduce essential mechanisms for the three main limitations in subgraph prediction: twin-tower module with the cross-attention mechanism for the collaborative limitation, joint prediction for the edge dependency limitation, and end-to-end learning for the global threshold limitation.

 \section{Experiments}
  We evaluate our models on two tasks: subgraph prediction and subgraph pattern prediction. Subgraph prediction focuses on general subgraphs, whereas  subgraph pattern prediction  focuses on the evolution of the predefined relationship between nodes in subgraphs. The experiments were conducted on a machine with Intel i7 8700K (CPU) and RTX2070.  We use Adam optimizer~\cite{adam} to train our model, and the initial learning rate is 0.005. In addition, all experiments were repeated  ten times, and the average performance of each method is reported.
  \subsection{Subgraph Prediction}
  \label{sec:SP}
    The subgraph prediction task evaluates the ability to capture the evolution of subgraphs in discrete-time dynamic graphs with multiple snapshots. Therefore, multiple snapshots are set as model inputs to predict the next snapshot in this task. As shown in Table~\ref{tab:datasets}, we use five public datasets from Network Repository~\cite{TNDR} for subgraph prediction. We split the dynamic graph into ten equal parts based on timestamps and construct ten snapshots for each dataset. The first nine snapshots are the training dataset, and the last snapshot is used for testing.

    The hyperparameters in our method are set as $D=128, k=10$, and $b=6$, which will be discussed in Sec.~\ref{sec:ana}.
    \begin{table*} 
        \center
        \caption{AUC scores of subgraph prediction  ($D=128, k=10, b=6$).}
      \label{tab:dlink}
      \begin{tabular}{rcccccccc}
          \toprule
          Dataset&\textbf{DynamicTriad}&\textbf{DynGEM}&\textbf{Dyngraph2vec}&\textbf{EvolveGCN}&\textbf{SPAN}&\textbf{Gain}\\
          \midrule
          ia-facebook  &0.833&0.753  &0.771&0.785&\textbf{0.921}&\textbf{10.56\%}\\
          soc-epinions &0.778&0.723  &0.791&0.804&\textbf{0.869}&\textbf{8.08\%}\\
          sx-askubuntu &0.816&0.811  &0.844&0.841&\textbf{0.908}&\textbf{7.58\%}\\
          sx-superuser &0.871&0.856  &0.884&0.896&\textbf{0.941}&\textbf{5.02\%}\\
          wiki-talk    &OOM &OOM &OOM&OOM&\textbf{0.943} &-\\
        \bottomrule
      \end{tabular}
      \end{table*} 
      \begin{table*}
        \center
        \caption{Parameter statistics of different methods ($D=128, k=10, b=6$).}
          \label{tab:params}
          \begin{tabular}{rccccccccc}     
            \toprule
            Dataset&\textbf{DynamicTriad}&\textbf{DynGEM}&\textbf{Dyngraph2vec}&\textbf{EvolveGCN}&\textbf{SPAN}\\
            \midrule
            ia-facebook  &15,776,259 &500,212,854   &72,184,652  &\textbf{61,527} &4,248,800 \\
            soc-epinions &44,294,931 &948,889,485   &M199,583,528  &26,175,567 &\textbf{9,001,464} \\
            sx-askubuntu &53,530,563 &1,067,104,722  &240,843,016   &18,225,525&\textbf{10,540,792} \\
            sx-superuser &65,212,947&1,213,544,964   &293,031,285  &80,957,922 &\textbf{12,487,856}\\
            wiki-talk    &OOM &OOM &OOM&OOM&\textbf{65,467,440} &-\\
          \bottomrule
        \end{tabular}
        \end{table*}  
    We select four state-of-the-art dynamic embedding methods as the baseline methods, including EvolveGCN~\cite{egcn}, Dyngraph2vec~\cite{dyngraph2vec}, DynamicTriad~\cite{triad}, and DynGEM~\cite{DynGEM}. For EvolveGCN, EvolveGCN-H is selected for comparison because it uses the GRU mechanism with better convergence. According to the suggestions in~\cite{egcn}, we specify the number of GCN layers as two and set the learning rate interval as [0.0001, 0.1]. For Dyngraph2vec, we use the AERNN version because it has higher accuracy than the default version. In addition, we set the look back size  as $l = {1, 2, 3}$ and other hyperparameters as suggested at~\cite{dyngraph2vec}. For DynamicTriad~\cite{triad}, we set hyperparameters $\beta1 \in  \{0.1, 1, 10\}$ and $\beta2 \in  \{0.1, 1, 10\}$ alternatively to achieve the best performance. For DynGEM, we set the initial sizes of autoencoders as $[500,300]$, $\alpha \in  [10^{-6},10^{-5}], \beta \in  [2, 5], v1 \in  [10^{-6},10^{-4}], v2 \in  [10^{-5},10^{-2}]$ as suggested in~\cite{DynGEM}.           
    We iteratively train SPAN via learning subgraph prediction in adjacent snapshots in chronological order, and other node embedding methods learn node embeddings for each snapshot.   
    In testing, SPAN predicts the subgraphs in $G_{10}$ based on $G_9$ directly, and other node embedding methods use the node embeddings of $G_9$ to predict links separately in these sampled subgraphs in $G_{10}$.

    Table~\ref{tab:dlink} lists the AUC scores of the subgraph prediction task. SPAN outperforms other state-of-the-art dynamic methods with an improvement from 5.02\% to 10.56\%. The method for calculating gain is the same as described in ~\cite{CTDNE}: (the accuracy of our method - the highest accuracy of other methods) / the highest accuracy of other methods. 
  According to Table~\ref{tab:dlink}, in our experiment, previous methods fail to predict subgraphs for large dynamic graphs (more than 1 million nodes) due to out-of-memory (OOM) in our experiment environment. As shown in Table~\ref{tab:params}, the number of  parameters in SPAN is lower than other methods and increases linearly with the graph size. The main reason is that SPAN can learn a subgraph with arbitrary size each time and is memory efficient compared to other methods based on the adjacency matrix of all snapshots or graphs. 
  \begin{table*}
    \center
      \caption{AUC scores of subgraph pattern prediction ($D=128, b=6$).}
      \label{tab:dsubgraph}
      \begin{tabular}{rcccccc}
        \toprule
        Dataset&\textbf{SPNN}&\textbf{HAN}&\textbf{HGT}&\textbf{SPAN}&\textbf{SPAN-H}&\textbf{Gain}\\
        \midrule
        email-eu &0.872 &0.859 &0.876 &0.885 &\textbf{0.941} &\textbf{7.42\%}\\
        DBLP &0.838&0.816&0.834 &0.861&\textbf{0.889} &\textbf{6.09\%}\\
        mathoverflow &0.855 &0.831 &0.840 &0.877&\textbf{0.915}&\textbf{7.01\%}\\
      \bottomrule
    \end{tabular}
    \end{table*}
  \subsection{Subgraph Pattern Prediction}
   \label{sec:SPP}
   Subgraph pattern prediction predicts that the subgraph would be transformed into a predefined subgraph pattern in a period of time. According to the task definition~\cite{SPNN}, we select three real-world heterogeneous dynamic graphs for this task. The email network Email-eu~\cite{email_euforevolution} is constructed using emails from a large European research institution, comprising 986 staff members, 50,572 emails, and 42 departments. The sampled subgraph has four staff members ($k=4$), and at least two staff work in different departments in the subgraph. The subgraph pattern is the communication between different departments, i.e., the connectivity between staff members in different departments. DBLP~\cite{DBLP} is a scientific paper co-author network with timestamps, including different entity types: 14,376 papers, 14,475 authors, 8,920 topics and 20 venues. We predict the evolution of 3-node subgraphs (author, topic, and venue), i.e., whether an author will publish in a venue and on a topic that the author has not published in the current snapshot. Mathoverflow~\cite{MITN} is a temporal network collected from the same website, including 24,818 users and 506,550 interactions, such as answers to questions, comments to questions, and comments to answers. The subgraph pattern is whether four users will interact more frequently over a period ($k=4$), i.e., whether there will be more edges of 4-node subgraphs in a period.

    The evolution that transforms the subgraph into a predefined subgraph pattern may involve establishing multiple edges at different times, implying that this evolution may include multiple intermediate states in multiple discrete snapshots.   
   However, we cannot ensure the appearance or nonappearance of subgraph patterns based on these intermediate states. Thus, to avoid these intermediate states, we discuss the appearance or nonappearance of subgraph patterns in a continuous-time rather than dividing the time into multiple snapshots. We divide each dynamic graph into two continuous-time graphs.  The first continuous-time graph is constructed using the first $70\%$ edges for training, and the second is constructed using the last $30\%$ edges for testing. We sample subgraphs from the first continuous-time graph and train the model by predicting  subgraph patterns in the first continuous-time graph. In testing, we predict subgraph patterns in the second continuous-time graph based on subgraphs in the first  continuous-time graph.

    We compare three state-of-the-art methods, namely, SPNN~\cite{SPNN}, HAN~\cite{HAN}, and HGT~\cite{HGT}, with SPAN-H in the subgraph pattern prediction task. SPNN is designed for subgraph pattern prediction based on limited and predefined subgraph patterns. HAN and HGT are based on the attention and graph convolution mechanisms to learn node embeddings for heterogeneous networks. We generate subgraph embeddings for HAN and HGT by averaging node embeddings and predicting subgraph pattern evolution using subgraph embeddings. We also set $D = 128$ and other hyperparameters, as suggested~\cite{SPNN,HAN,HGT}. Table~\ref{tab:dsubgraph} shows the AUC scores of subgraph pattern prediction on three heterogeneous networks, and that SPAN-H achieves the best performance increasing from 6.09\% to 7.42\%, compared to other methods. The proposed model can learn the evolution of subgraphs in heterogeneous dynamic graphs more effectively.  Compared with previous methods, SPAN-H completely learns the existing subgraph patterns from  data and uses a twin-tower module with the cross-attention mechanism to capture the evolution of subgraphs based on diverse information.

     \subsection{Model Analysis}
     \label{sec:model}  
  
\begin{figure*}
        \centering 
        \begin{subfigure}{.13\textwidth}
        \includegraphics[width=1\linewidth,height=4.5cm]{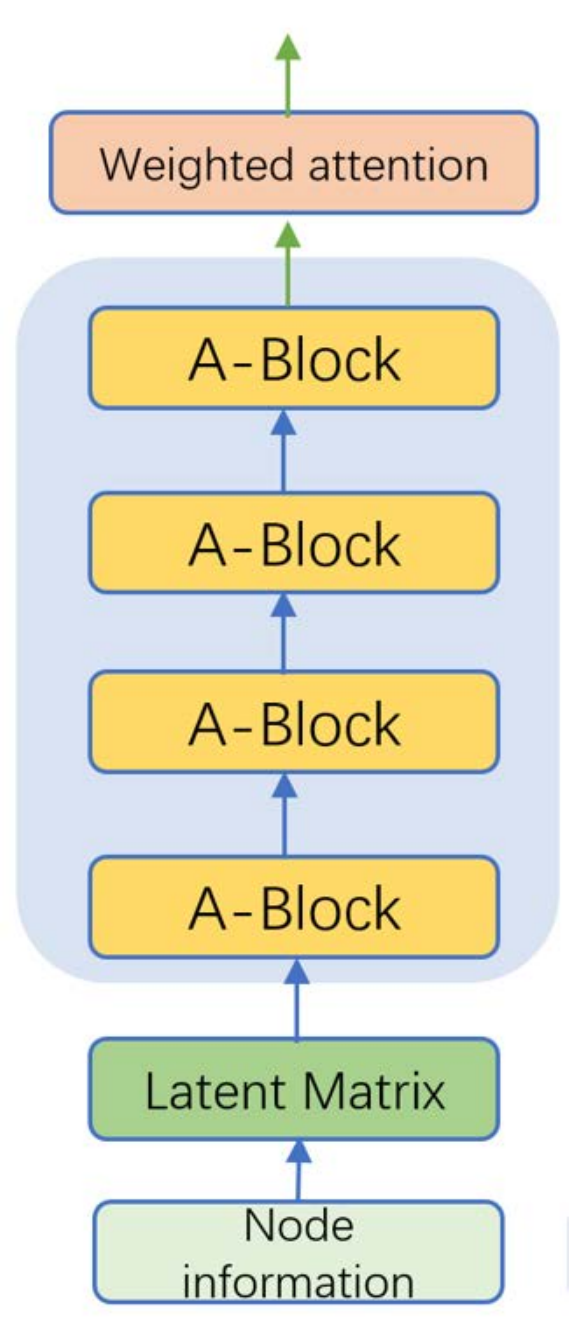}  
        \caption{model 1}
        \label{fig:model_1}
        \end{subfigure}
        \begin{subfigure}{.18\textwidth}
          \includegraphics[width=1\linewidth,height=4.5cm]{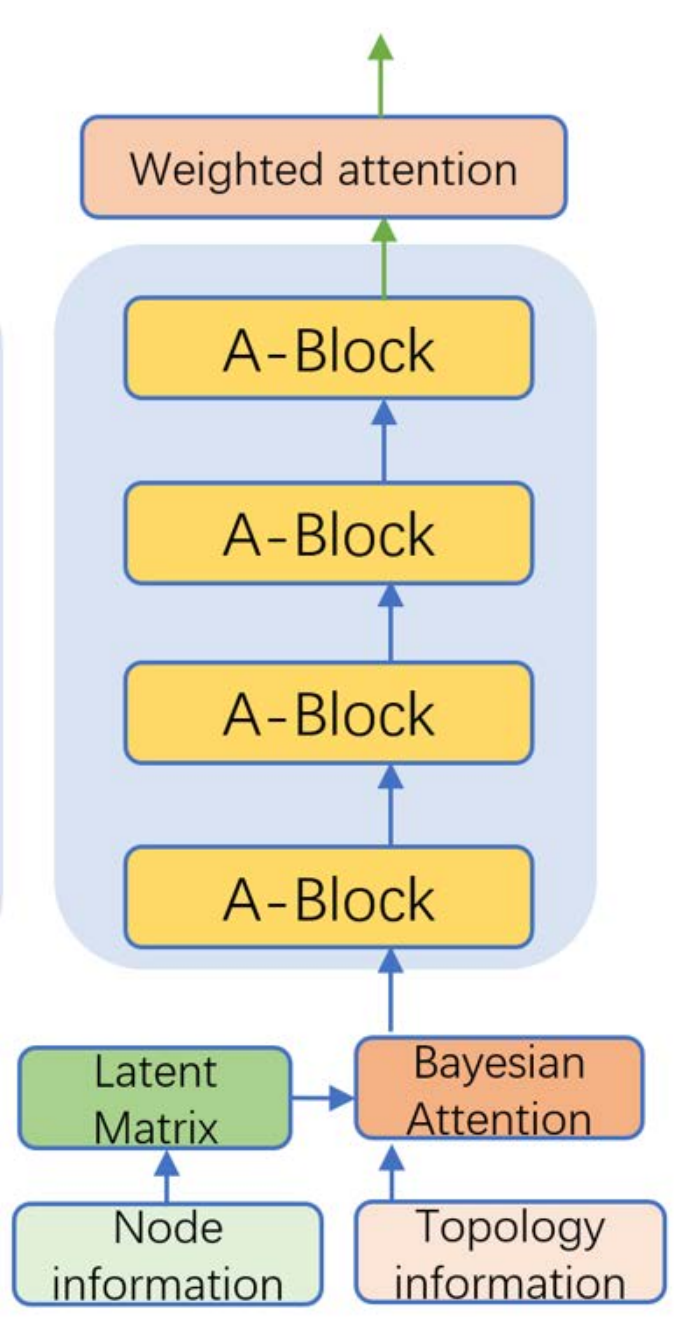}  
          \caption{model 2}
          \label{fig:model_2}
        \end{subfigure} 
        \begin{subfigure}{.25\textwidth}
          \includegraphics[width=1\linewidth,height=4.5cm]{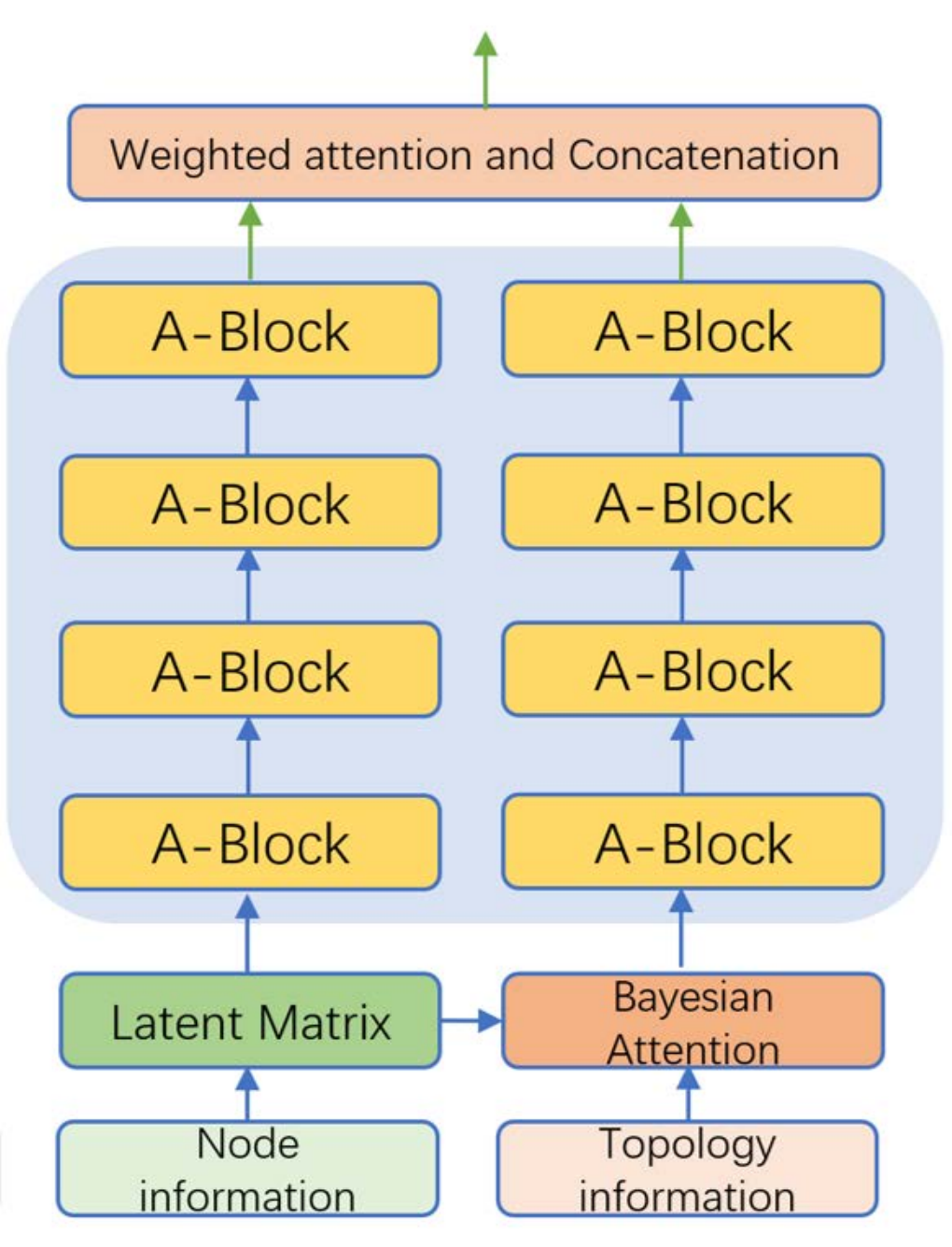}  
          \caption{model 3}
          \label{fig:model_3}
        \end{subfigure}
        \begin{subfigure}{.26\textwidth}
          \includegraphics[width=1\linewidth,height=4.5cm]{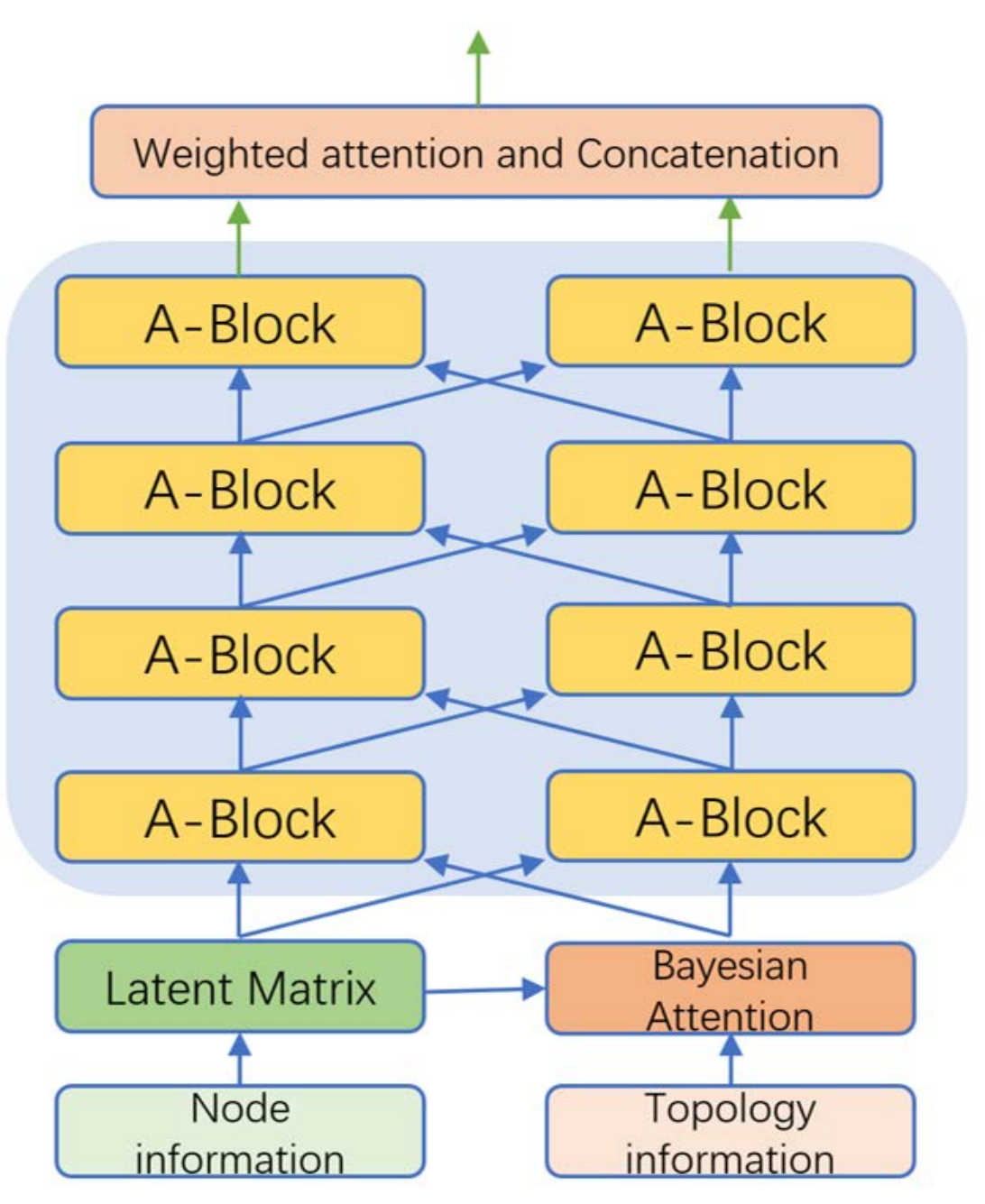}  
          \caption{model 4}
          \label{fig:model_4}
        \end{subfigure}
        \caption{Model architectures for comparison: (a) model 1 using only node attribute information,
          (b)  model 2 using only context information, 
          (c) model 3 using both context information and node attribute information,
          (d) model 4 using a cross-attention mechanism to fuse node and context information in multi-level.}
        \label{fig:structure}
      \end{figure*}

\begin{figure*}
        \centering
        \begin{subfigure}{.4\textwidth}
          \includegraphics[width=5cm, height=4.5cm]{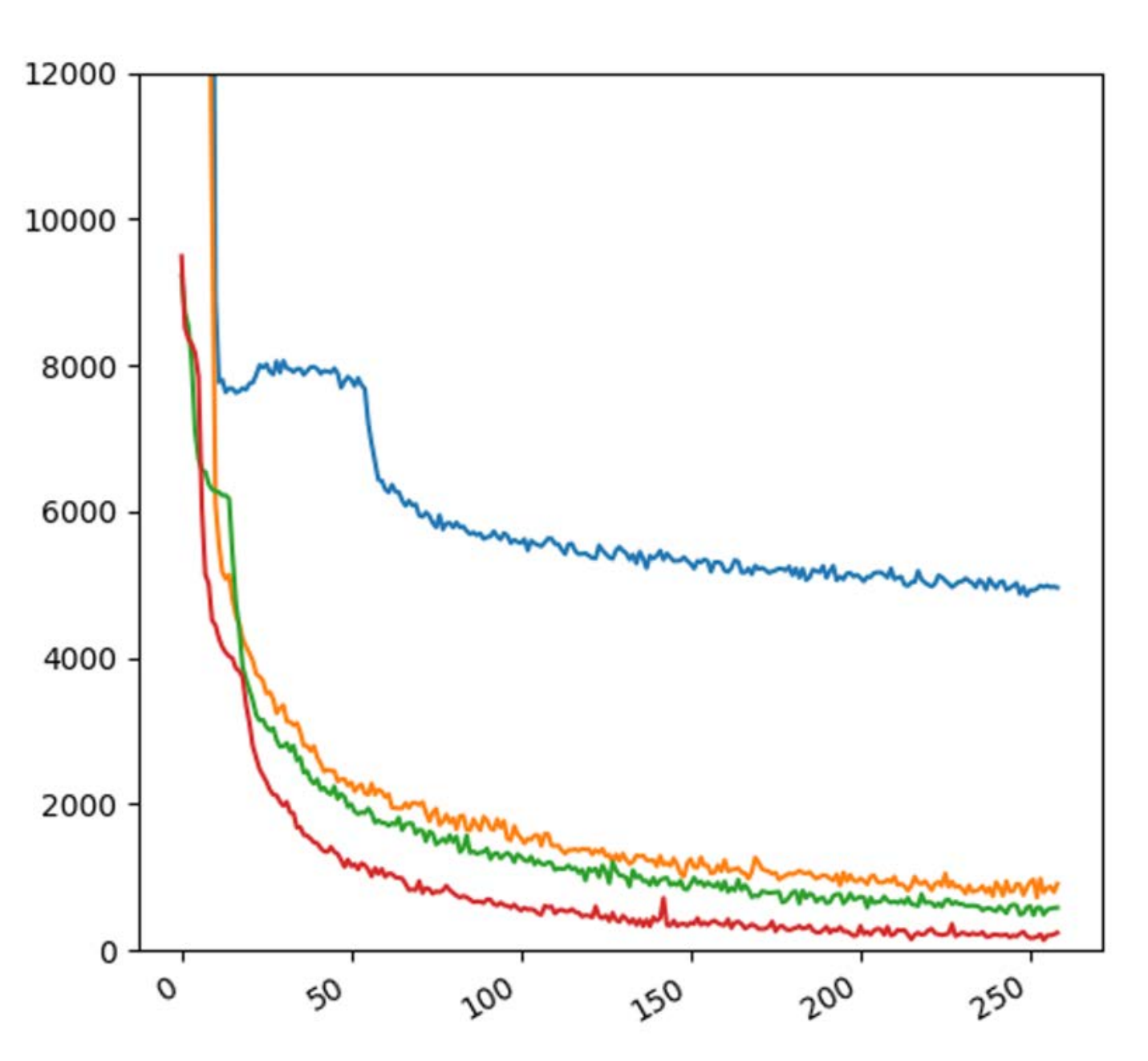}  
          \caption{Losses of four models}
          \label{fig:loss}
        \end{subfigure}
        \begin{subfigure}{.41\textwidth}
          \includegraphics[width=5cm, height=4.5cm]{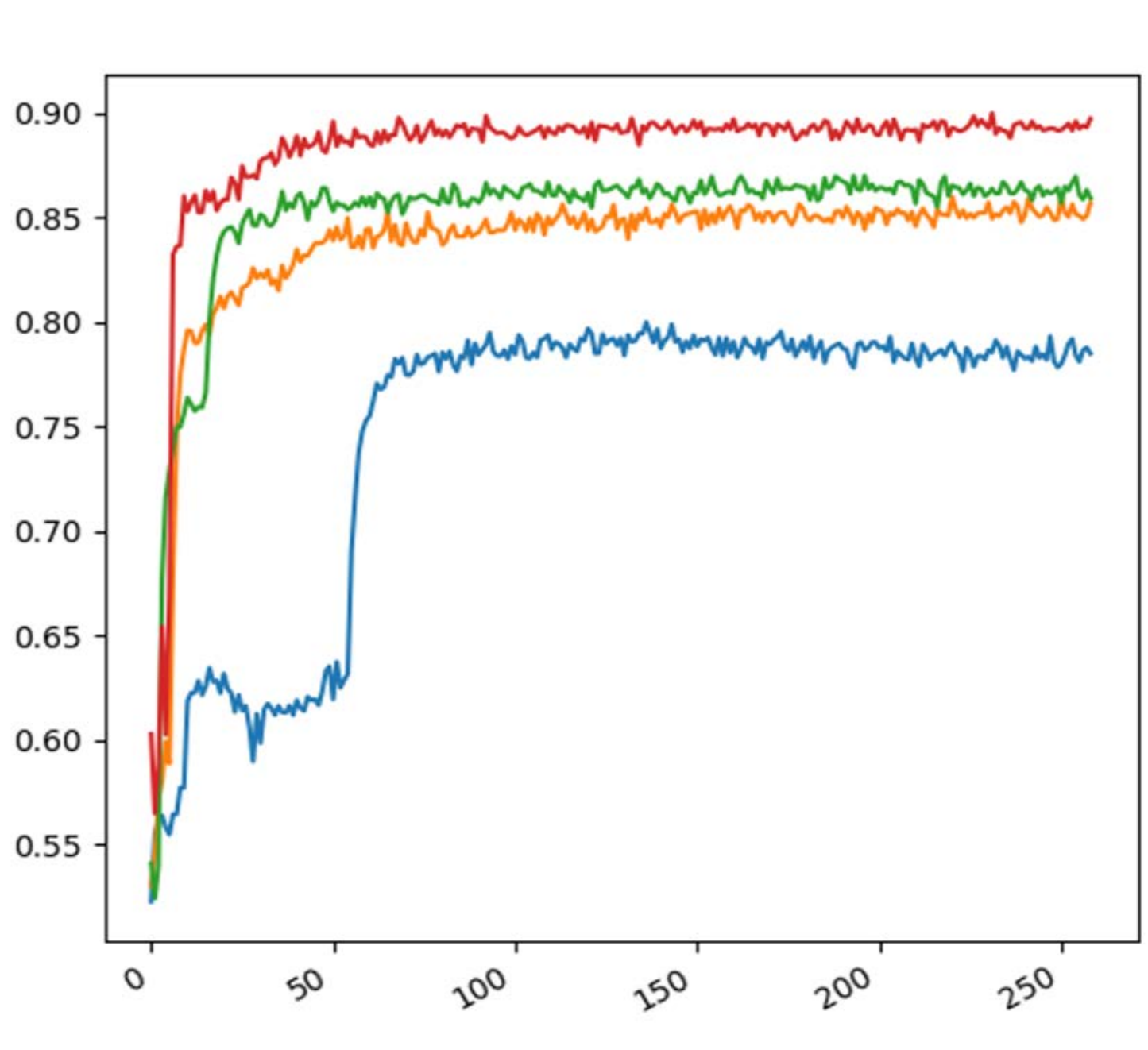}  
          \caption{AUC scores of four models}
          \label{fig:acc}
        \end{subfigure}
        \begin{subfigure}{.1\textwidth}
          \includegraphics[width=1.2cm, height=4.5cm]{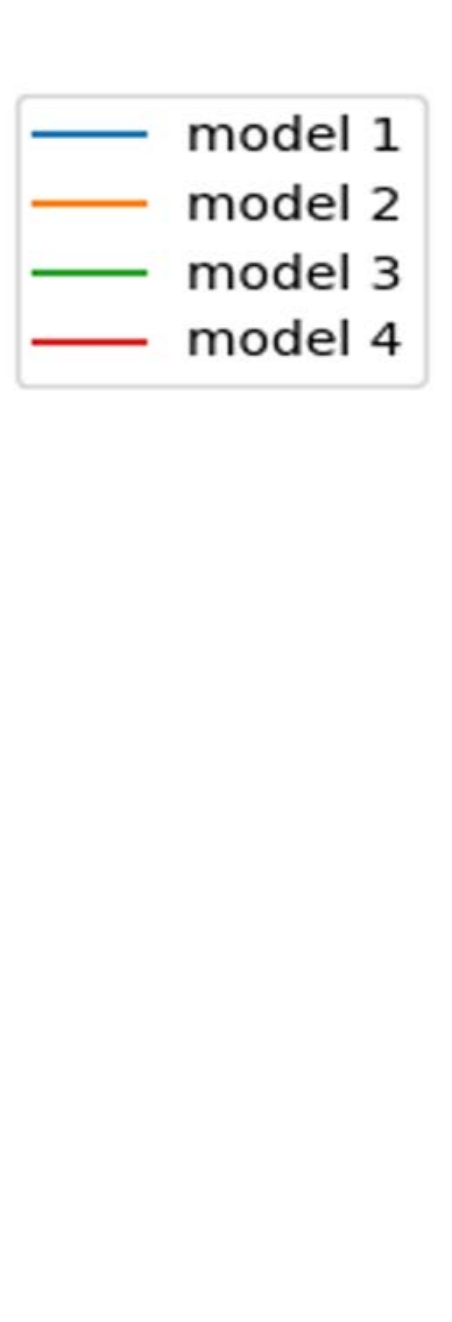}  
        \end{subfigure}
        \caption{ 
          The loss curves (a) and AUC score curves (b) of four models over epochs. These curves are averaged across three datasets in subgraph pattern prediction.
           }
        \label{fig:loss_and_acc}
      \end{figure*}

     Our model achieves the  best performance on both subgraph prediction and subgraph pattern prediction tasks. We attribute this benefit to some new mechanisms, such as Bayesian attention, the twin-tower module and the cross-attention mechanism. 
     In this section, we discuss how these mechanisms gradually improve our model. As shown in Figure~\ref{fig:structure}, we design four models for comparative analysis on the subgraph pattern prediction task using the same dataset in Sec.~\ref{sec:SPP}. 
     As shown in  Figure~\ref{fig:structure}, model 1 only uses node attribute information, model 2 uses context information (the fusion of node attribute information and topology information generated by Bayesian attention), model 3 extends the tower module based on model 2, and model 4 extends  model 3 with cross-attention. The reason for choosing  this task is that heterogeneous networks have rich subgraph patterns and have numerous practical applications. Figure~\ref{fig:loss_and_acc} shows the results of the four models.   
     
     As shown in Figure~\ref{fig:loss_and_acc}, model 2 outperforms model 1, which means the node attribute information and topology information both play a significant role in subgraph evolution.
      \begin{figure*}
        \center
        \begin{subfigure}[p]{0.22\textwidth}
        \resizebox{\linewidth}{!}{
        \begin{tikzpicture}
        \begin{axis}[title=, legend pos=outer north east, xlabel=$k$]
        \addplot[color=blue, mark=triangle]
        table[row sep=crcr]
        {
        X Y \\
        4 0.943 \\
        7 0.935 \\
        10 0.931 \\
        13 0.917 \\
        16 0.886 \\
        };
        \addplot[color=black, mark=x]
        table[row sep=crcr]
        {
        X Y \\
        4 0.868 \\
        7 0.869 \\
        10 0.866 \\
        13 0.852 \\
        16 0.833 \\
        };
        \addplot[color=green, mark=square]
        table[row sep=crcr]
        {
        X Y \\
        4 0.918 \\
        7 0.911 \\
        10 0.906 \\
        13 0.902 \\
        16 0.896 \\
        };
        \addplot[color=olive, mark=o]
        table[row sep=crcr]
        {
        X Y \\
        4 0.953 \\
        7 0.944 \\
        10 0.937 \\
        13 0.928 \\
        16 0.916 \\
        };
        \addplot[color=red, mark=star]
        table[row sep=crcr]
        {
        X Y \\
        4 0.948 \\
        7 0.946 \\
        10 0.943 \\
        13 0.937 \\
        16 0.923 \\
        };
        \end{axis}
        \end{tikzpicture}
        }
        \caption{Subgraph size $k$}   
        \label{fig:subc}
        \end{subfigure}
        \begin{subfigure}[p]{0.22\textwidth}
        \resizebox{\linewidth}{!}{
        \begin{tikzpicture}
        \begin{axis}[title=, legend pos=outer north east, xlabel=$k$]
        \addplot[color=blue, mark=triangle]
        table[row sep=crcr]
        {
        X Y \\
        4 18 \\
        7 21 \\
        10 37 \\
        13 81 \\
        16 135 \\
        };
        \addplot[color=black, mark=x]
        table[row sep=crcr]
        {
        X Y \\
        4 20 \\
        7 26 \\
        10 40 \\
        13 90 \\
        16 158 \\
        };
        \addplot[color=green, mark=square]
        table[row sep=crcr]
        {
        X Y \\
        4 22 \\
        7 27 \\
        10 43 \\
        13 96 \\
        16 164 \\
        };
        \addplot[color=olive, mark=o]
        table[row sep=crcr]
        {
        X Y \\
        4 32 \\
        7 38 \\
        10 56 \\
        13 112 \\
        16 192 \\
        };
        \addplot[color=red, mark=star]
        table[row sep=crcr]
        {
        X Y \\
        4 142 \\
        7 167 \\
        10 202 \\
        13 251 \\
        16 347 \\
        };
        \end{axis}
        \end{tikzpicture}
        }
        \caption{Runtime}
        \label{fig:subd}
        \end{subfigure}
        \begin{subfigure}[p]{0.22\textwidth}
        \resizebox{\linewidth}{!}{
        \begin{tikzpicture}
        \begin{axis}[title=, legend pos=outer north east, xlabel=$q$]
        \addplot[color=blue, mark=triangle]
        table[row sep=crcr]
        {
        X Y \\
        16 0.868 \\
        32 0.902 \\
        64 0.926 \\
        128 0.929 \\
        256 0.931 \\
        };
        \addplot[color=black, mark=x]
        table[row sep=crcr]
        {
        X Y \\
        16 0.809 \\
        32 0.832 \\
        64 0.857 \\
        128 0.861 \\
        256 0.866 \\
        };
        \addplot[color=green, mark=square]
        table[row sep=crcr]
        {
        X Y \\
        16 0.843 \\
        32 0.876 \\
        64 0.901 \\
        128 0.905 \\
        256 0.906 \\
        };
        \addplot[color=olive, mark=o]
        table[row sep=crcr]
        {
        X Y \\
        16 0.872 \\
        32 0.911 \\
        64 0.933 \\
        128 0.934 \\
        256 0.937 \\
        };
        \addplot[color=red, mark=star]
        table[row sep=crcr]
        {
        X Y \\
        16 0.882 \\
        32 0.913 \\
        64 0.936 \\
        128 0.941 \\
        256 0.943 \\
        };
        \end{axis}
        \end{tikzpicture}
        }
        \caption{Dimension $D$}
        \label{fig:suba}
        \end{subfigure}
        \begin{subfigure}[p]{0.3\textwidth}
        \resizebox{\linewidth}{!}{
        \begin{tikzpicture}
        \begin{axis}[title=, legend pos=outer north east, xlabel=$b$]
        \addplot[color=blue, mark=triangle]
        table[row sep=crcr]
        {
        X Y \\
        2 0.921 \\
        4 0.925 \\
        6 0.931 \\
        8 0.933 \\
        10 0.934 \\
        };
        \addplot[color=black, mark=x]
        table[row sep=crcr]
        {
        X Y \\
        2 0.849 \\
        4 0.852 \\
        6 0.866 \\
        8 0.868 \\
        10 0.871 \\
        };
        \addplot[color=green, mark=square]
        table[row sep=crcr]
        {
        X Y \\
        2 0.898 \\
        4 0.901 \\
        6 0.906 \\
        8 0.906 \\
        10 0.908 \\
        };
        \addplot[color=olive, mark=o]
        table[row sep=crcr]
        {
        X Y \\
        2 0.927 \\
        4 0.933 \\
        6 0.937 \\
        8 0.939 \\
        10 0.942 \\
        };
        \addplot[color=red, mark=star]
        table[row sep=crcr]
        {
        X Y \\
        2 0.931 \\
        4 0.938 \\
        6 0.943 \\
        8 0.944 \\
        10 0.946 \\
        };
        \addlegendentry{ia-facebook}
        \addlegendentry{soc-epinions}
        \addlegendentry{sx-askubuntu}
        \addlegendentry{sx-superuser}
        \addlegendentry{wiki-talk}
        \end{axis}
        \end{tikzpicture}
        }
        \caption{Blocks $b$}   
        \label{fig:subb}
        \end{subfigure}
        \caption{(a), (c), and (d) are the AUC scores of discrete-time subgraph prediction for different settings on hyperparameters $k$, $D$ and $b$; (b) is the training time of SPAN with different $k$ (minutes). } 
        \label{fig:group}
        \end{figure*}
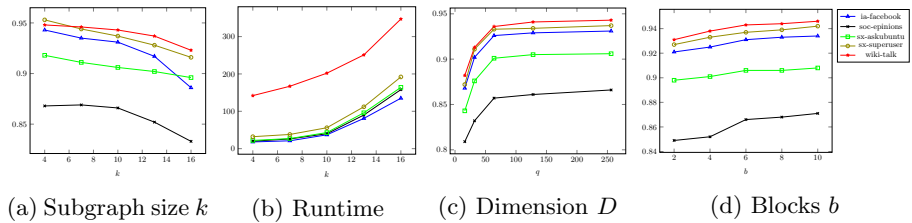 
     In addition, model 3 outperforms model 2 because node attribute information will become fuzzy during the Bayesian attention process in model 2 even though context information can be regarded as the fusion of node attribute information and topology information.
    Figure~\ref{fig:loss_and_acc} also shows that model 4 achieves higher accuracy and faster convergence speed than model 3. The benefit is attributable to the cross-attention mechanism. Important topology information or node attribute information may also be related to each other in subgraph evolution.
    The cross-attention mechanism integrates topology information and node attribute information to extract important features rather than isolated extraction, enabling our method to avoid feature loss. 
  
     \subsection{Hyperparameter Analysis}
     \label{sec:ana} 
       We discuss the significant hyperparameters of subgraph sampling and model learning,  respectively.
      \textbf{For subgraph sampling}, the maximum node size of  subgraphs is $k$ even though subgraphs can be generated with an arbitrary size ($\le k$) in the subgraph sampling process. Figure~\ref{fig:group}(a) illustrates that smaller subgraphs can be learned better than larger subgraphs because the evolution pace of bigger subgraphs becomes larger, and effectively predicting these subgraphs is difficult. The number of nodes also affects the training time. We set the hyperparameters ($D=128$ and $b=6$) to evaluate the training time of SPAN. Figure~\ref{fig:group}(b) shows that the training time increases with k and is almost linear with the subgraph size.
      \textbf{For model learning}, the dimension of node embedding vectors in the latent matrix is $D$, and Figure~\ref{fig:group}(c) demonstrates performance changes with different $D$.  Our model's number of attention-based blocks is $b$, and the stacking of multiple blocks is used to learn the complex function for subgraph evolution.  Figure~\ref{fig:group}(d) shows that the model's performance is generally proportional to $b$, as fewer blocks would be under-fitted. Since the model's performance increases gradually when $b>6$, we set the hyperparameter $b$ = 6 to reduce the number of parameters and avoid overfitting.

     \section{Conclusion}
     \label{sec:con}
     
     In this study, we propose a novel end-to-end model for subgraph prediction in dynamic graphs. We evaluate our model by comparing it with several state-of-the-art methods, including node embedding-based methods and graph neural network-based methods in dynamic graphs with two tasks: subgraph prediction and subgraph pattern prediction. Experimental results demonstrate that our model can achieve substantial gains, from 5.02\% to 10.88\%.
     \section{Acknowledgements}
     \label{sec:ack}
     This work was supported by National Natural Science Foundation of China (61972343).
        
\bibliographystyle{splncs04}
\bibliography{sample-base}

\end{document}